\documentclass[12pt]{article}
\usepackage{amsmath}
\usepackage{amssymb}
\usepackage{amscd}
\usepackage{amsfonts}
\usepackage{amsthm}
\usepackage{mathrsfs}
\usepackage[matrix,arrow,curve]{xy}
\usepackage{amsbsy}
\usepackage{slashed}
\usepackage{amssymb}
\usepackage{bbm}
\makeindex
  \newtheorem{theorem}{Theorem}[section]
  \newtheorem{prop}[theorem]{Proposition}
  \newtheorem{lemma}[theorem]{Lemma}
  
\theoremstyle{remark}
  \newtheorem{ex}{Example}[section]
  \newtheorem{rem}[theorem]{Remark}

 \newcounter{biscompt}

 \theoremstyle{plain}

\usepackage{hyperref}

 \numberwithin{equation}{section}

\newcommand{\Dj}{\hbox to 8pt{\raisebox{.4\height}{-}\hss D}}
\newcommand{\id}{\ensuremath{{\rm id}}}

\newcommand{\lc}{\ensuremath{\mathcal L}}

\newcommand{\tc}{\ensuremath{\mathcal T}}

\newcommand{\lt}{\langle}
\newcommand{\rt}{\rangle}

\newcommand{\beq}[1]{\begin{equation}\label{#1}}
\newcommand\eeq{\end{equation}}
\newcommand\bqa {\begin{eqnarray}}
\newcommand\eqa {\end{eqnarray}}

\newcommand{\bear}{\begin{array}}
\newcommand{\enar}{\end{array}}

\newcommand{\R}{\mathbb{R}}

\newcommand{\slnr}{\ensuremath{\mathfrak{sl}_n(\mathbb R)}}
\newcommand{\glnr}{\ensuremath{\mathfrak{gl}_n(\mathbb R)}}
\newcommand{\sonr}{\ensuremath{\mathfrak{so}_n(\mathbb R)}}
\newcommand{\sof}{\ensuremath{\mathfrak{so}_4(\mathbb R)}}
\newcommand{\bp}{\ensuremath{\mathfrak{b}^+_n(\mathbb R)}}
\newcommand{\bm}{\ensuremath{\mathfrak{b}^-_n(\mathbb R)}}
\newcommand{\bmf}{\ensuremath{\mathfrak{b}^-_4(\mathbb R)}}
\newcommand{\Slnr}{\ensuremath{SL_n(\mathbb R)}}
\newcommand{\Sonr}{\ensuremath{SO_n(\mathbb R)}}
\newcommand{\Bp}{\ensuremath{B^+_n(\mathbb R)}}
\newcommand{\Bm}{\ensuremath{B^-_n(\mathbb R)}}
\newcommand{\sy}{\ensuremath{Symm_n(\mathbb R)}}

\newcommand{\dd}{\partial}

\newcommand{\tr}{\mathrm{tr}\,}

\usepackage[usenames]{color}

\usepackage{tikz}

\usepackage{graphicx}

\begin{document}
\begin{center}
\Large{\bf{Full symmetric Toda system and vector fields on the group $SO_n(\R)$}}

\ \\

Yu.B. Chernyakov\footnote{National Research Center "Kurchatov Institute"\ , Moscow, Russia.}$^{,}$\footnote{Moscow Institute of Physics and Technology, Dolgoprudny, Moscow region, Russia}$^{,}$\footnote{Institute for Information Transmission Problems of the Russian Academy of Sciences (Kharkevich Institute), Moscow, Russia}, chernyakov@itep.ru\\
G.I. Sharygin\footnotemark[1]$^{,}$\footnotemark[2]$^{,}$\footnote{Moscow State University, Faculty of Mechanics and Mathematics, Leninskie gory d. 1, Moscow, 119991, Russia}$^{,}$\footnote{Sino-Russian Mathematics Center, PKU Beijing, PRC}, sharygin@itep.ru\\

\end{center}
\begin{abstract}
In this paper we discuss the relation between the functions that give first integrals of full symmetric Toda system (an important Hamilton system on the space of traceless real symmetric matrices) and the vector fields on the group of orthogonal matrices: it is known that this system is equivalent to an ordinary differential equation on the orthogonal group, and we extend this observation further to its first integrals. As a by-product we describe a representation of the Lie algebra of $B^+(\R)$-invariant functions on the dual space of Lie algebra $\mathfrak{sl}_n(\R)$ (under the canonical Poisson structure) by vector fields on $SO_n(\R)$.
\end{abstract}

\section{Introduction}
Recall that Toda systems are a broad class of integrable systems, that are inspired by (or just close to) the famous \textit{open Toda chain}, the integrable system, that describes the behaviour of an $n$-tuple of material points on a straight line. If $q_1,\dots,q|_n$ are the coordinates of these points and $p_1,\dots,p_n$ are their momenta, we can introduce the dynamics simply by considering the suitable Hamilton function: then the standard Poisson structure
\[
\{p_i,q_j\}=\delta_{ij}
\]
will take care of the rest. In the theory of the open Toda chain the choice of Hamilton function is prescribed by the formula
\[
H(p,q)=\sum_{i=1}^n\frac12p_i^2+\sum_{j=1}^{n-1}e^{q_j-q_{j+1}}.
\]
Properties of the open Toda chain have been investigated by numerous authors in 1960-ies and later. The crucial step in these investigations was made by Flaschka, who discovered the Lax form of this system: if we make a change of variables: $a_i=\frac{1}{\sqrt 2}p_i,\,b_j=\frac{1}{2}e^{\frac12(q_j-q_{j+1})}$, then the Toda system will take the following Lax form:
\beq{toda0}
\dot L=[M(L),L],
\eeq
where $L,\,M(L)$ are the matrices
\[
L=\begin{pmatrix}
       a_1 & b_1 & 0 & 0 & \dots & 0\\
       b_1 & a_2 & b_2 & 0 & \dots & 0\\
       0    & b_2 & a_3 & b_3& \dots & 0\\
       \vdots & & \ddots &\ddots & \ddots & \vdots\\
       0 & & \dots & b_{n-2} & a_{n-1} & b_{n-1}\\
       0 & & \dots &       0     & b_{n-1} & a_n
    \end{pmatrix},
    M(L)=\begin{pmatrix}
       0 & b_1 & 0 & 0 & \dots & 0\\
       -b_1 & 0 & b_2 & 0 & \dots & 0\\
       0    & -b_2 & 0 & b_3& \dots & 0\\
       \vdots & & \ddots &\ddots & \ddots & \vdots\\
       0 & & \dots & -b_{n-2} & 0 & b_{n-1}\\
       0 & & \dots &       0     & -b_{n-1} & 0
    \end{pmatrix}.
\]
This equation, or rather its generalisations will be in the center of our attention in this paper.

Namely, the \textit{full symmetric Toda system} now can be described as a pretty na\"\i ve generalisation of the usual (open) Toda chain: just replace the tri-diagonal symmetric matrix $L$ in equation \eqref{toda0} by an arbitrary symmetric matrix, and use the same na\"\i ve anti-symmetrization procedure, as before. That's, take the matrices $L,\,M(L)$ as
\[
L=\begin{pmatrix}
       a_{11} & a_{12} & a_{13} &  \dots & a_{1n}\\
       a_{12} & a_{22} & a_{23} &  \dots & a_{2n}\\
       a_{13}  & a_{23} & a_{33} & \dots & a_{3n}\\
       \vdots &\vdots &  & \ddots & \vdots\\
       a_{1n} & a_{2n} & \dots      & a_{n-1,n} & a_{nn}
    \end{pmatrix},
    M(L)=\begin{pmatrix}
       0 & a_{12} & a_{13} &  \dots & a_{1n}\\
       -a_{12} & 0 & a_{23} &  \dots & a_{2n}\\
       -a_{13}  & -a_{23} & 0 & \dots & a_{3n}\\
       \vdots &\vdots &  & \ddots & \vdots\\
       -a_{1n} & -a_{2n} & \dots & -a_{n-1,n} & 0
    \end{pmatrix}.
\]
Then the equation \eqref{toda0} for this choice of variables is called ``full symmetric Toda system''; it allows lots of further generalizations and numerous variations, beginning with considering analogs of symmetric matrix spaces related with Cartan decompositions of real semisimple Lie algebras, i.e. the spaces $\mathfrak p$ in the orthogonal decompositions
\[
\mathfrak g=\mathfrak k\oplus\mathfrak p.
\]
One can go further all the way down to choosing suitable subspaces in the space of symmetric matrices, the only condition being that commutators with the na\"\i ve antisymmetrization map preserve the structure of zeros in the matrix $L$. The behaviour of trajectories of this system is closely related with the topology of homogenous spaces associated with the real Lie groups (flag spaces and their generalisations), and has been in the focus of attention of numerous researchers throughout the last quarter of the century at least.

The main purpose of this paper is to bring together a series of statements and formulas, that seem to have always been at the periphery of many papers, dealing with the full symmetric Toda system, but (up to the knowledge of the authors) have never been explicitly formulated and proved, to say nothing about being put together in a common paper (see however the review by Olshanetsky and Perelomov, \cite{OP} where such constructions are also duscussed); here we limit ourselves to the ``classical'' situation of usual symmetric matrices, although all the constructions can without any effort be transferred to the generalised case i.e. to the space of ``generalised symmetric matrices'' associated with arbitrary real semisimple Lie algebra. These statements include a very explicit description of the Poisson structure on the space of symmetric matrices, phrased in terms of ``symmetric'' and ``usual'' matrix-gradients of a function (formulas \eqref{pois2.1}, \eqref{pois3}, \eqref{pois2.2} and \eqref{pois4}), the formulas for the Poisson bracket of \Bp-invariant functions and their Hamilton vector fields: proposition \ref{instrumental}, and remarks \ref{rem1imp} and \ref{remhamfield}. Observe that the statement of proposition \ref{instrumental} is a certain specification and slight generalization of the famous AKS principle, which allows one to produce commuting integrals from invariant functions: here we obtain a full anti-homomorphism of Lie algebras, which gives the usual commuting elements, when restricted to commutative subalgebras. Finally, we also give a relation between the \Bp-invariant functions on \slnr\ and the vector fields on the group \Sonr: we get an explicit homomorphism from the algebra of \Bp-invariant functions on \slnr\ (with respect to the Poisson bracket on \slnr) and the algebra of vector fields on \Sonr.

Among other things that follow directly from the formulas we present here are the \textit{superintegrability} of the full symmetric Toda system (it follows from the description of the algebra of \Bp-invariant functions; also we obtain a method to construct commutative families of first integrals of the systems, and symmetries of the corresponding system on the group \Sonr. All these statements seem to have been known (especially the superintegrability, which appeared in various papers before, e.g. in \cite{CS2} or \cite{RSchr}), but the methods of proof and the explicitness and generality of the formulas we use here, in our humble opinion deserve special attention.

Structure of the paper is very simple: in the section \ref{sect2} we describe the Poisson structure on the space of symmetric matrices, which we introduce from the isomorphism with the space $\bm^*$, induced by a nondegenerate pairing $\sy\otimes\bm\to\R$. In particular, in section \ref{sect2.2} we describe the Poisson pairing of \Bp-invariant functions; after this in sections \ref{sect3.2} and \ref{sect3.3} we construct the vector fields on the group \Sonr\ from \Bp-invariant functions and show that this construction is in fact a homomorphism of Lie algebras. Finally, in the end of our paper we calculate an example of such functions on $Symm_4(\mathbb R)$ and the corresponding vector fields and describe their commutation relations: something that has long baffled one of the authors of the present paper.

\paragraph{Important warning!} For our purposes it is convenient to consider functions and vector fields on Euclidean spaces and matrix groups that are analogs of meromorphic functions and vector fields in analytic geometry; speaking a bit loosely, such functions should be equal to ratios of usual smooth functions (if the denominator vanishes in a proper submanifold of positive codimension) and vector fields should have coefficients of this form. We are not going to deal with the general theory of such rational functions, which should be quite hard since the algebra of smooth functions is not finitely generated, has zero divisors and other pathological properties. However in every particular the functions that we would use will be of that sort. In particular, they are ``smooth in generic point'' and have ``poles'' at certain spaces. To save the time and notation we will denote such functions simply by $C^\infty(\cdot)$.

\section{Poisson structure on the space \sy}\label{sect2}
\subsection{The general construction}
Recall that \slnr\ is the Lie algebra of traceless matrices with real entries, that matrix commutator playing the role of Lie bracket. Below we will denote by $e_{ij},\,i,j=1,\dots,n$ ($i$ is the number of row, and $j$ the number of column, where the unit stands) the standard basis of matrix units in $Mat_n(\R)$, then the basis in \slnr\ can be taken in the form $\{e_{ij},i\neq j, e_{ii}-e_{nn}\}$. Below we will not distinguish between the elements of \slnr\ and those of the general matrix Lie algebra $\mathfrak{gl}_n(R)$, so we'll denote the generic element of either algebra by
\[
x=\sum_{i,j=1}^nx_{ij}e_{ij}
\]
mutely assuming that $\sum_{i=1}^nx_{ii}=0$ in the case of \slnr. The commutator relation in both cases is given by the formula
\[
[e_{ij},e_{pq}]=\delta_{pj}e_{iq}-\delta_{iq}e_{pj}.
\]
Let $f,g \in C^\infty(\slnr^*)$ be any two smooth functions; recall that the Lie-Poisson bracket on the algebra $C^\infty(\slnr^*)$ is given by the formula
\beq{pois1}
\{f,g\}(x)=(x,[df(x),dg(x)]),
\eeq
where $(,)$ is the natural pairing between $\slnr$ and its dual space $\slnr^*$ and $[,]$ is the commutator; here and below we will denote by $df(x)$ the \textit{usual} differential of a smooth function $f$ on $\mathfrak{sl}_n(\R)^*$ at the point $x\in\mathfrak{sl}_n(\R)^*$, which we regard as the linear map $df(x):\mathfrak{sl}_n(\R)^*\cong T_x\mathfrak{sl}^*_n(\R)\to\R$; as such $df(x)\in\slnr$. The same formula works in case of an arbitrary Lie algebra.

The ``Killing pairing'' on \glnr\ and $\slnr\subset\glnr$ is given by the formula
\beq{kill1}
\lt A,B\rt=\tr(A^tB),
\eeq
for any two matrices $A,B$ (here $A^t$ denotes the transposed matrix), so that:
\beq{kill2}
\lt A,B\rt=\sum_{i,j=1}^na_{ij}b_{ij},\ \mbox{if}\ A=\sum_{i,j=1}^n a_{ij}e_{ij},\ B=\sum_{i,j=1}^n a_{ij}e_{ij}.
\eeq
This pairing looks like the canonical Euclidean pairing in the basis $e_{ij}$ and hence is clearly non-degenerate, so it can be used to identify $\slnr^*\cong\slnr$, which consists simply of ``raising the indices''. In terms of this identification we can regard $x$ as an element of \slnr\ in the formula \eqref{pois1}; then we replace the differentials $df(x)$ in it by ``matrix gradients'' $\nabla f(x),\,\nabla g(x)$ of the functions $f,g\in C^\infty(\slnr)$ and change the natural pairing on dual spaces by the Killing pairing. So eventually we get
\beq{pois1.5}
\{f,g\}(x)=\lt x,[\nabla f(x),\nabla g(x)]\rt,
\eeq
for all $f,g\in C^\infty(\slnr),\,x\in\slnr$, where
\[
\nabla f(x)=\sum\frac{\dd f}{\dd x_{ij}}e_{ij}
\]
denotes the matrix of partial derivatives of $f$.
\begin{rem}
Strictly speaking the Lie theoretic Killing form on \glnr\ and \slnr\ is different from the pairing \eqref{kill1}: by definition Killing form on a Lie algebra is given by the formula $K(A,B)=\tr(ad(A)ad(B))$ where $ad$ denotes the adjoint representation
\[
ad(X)(Y)=[X,Y].
\]
In particular in terms of the matrices $A,B$ we have
\[
K(A,B)=\tr(ad(A)ad(B))=2n\tr(AB)-2\tr(A)\tr(B).
\]
As one sees apart from the absence of the transposition we also need the scaling factor $2n$ in the case of \slnr. However the formulas \eqref{kill1}, \eqref{kill2} are traditionally used in the study of symmetric Toda system on \sy, and we will also stick to that terminology.
\end{rem}
It is clear that the restriction of the pairing \eqref{kill2} to the subspaces $Symm_n(\R)\subset\slnr$ of real symmetric traceless matrices and $\bm\subset\slnr$ (viewed as the space of the traceless upper-triangular matrices) gives a non-degenerate pairing $\lt,\rt: Symm_n(\R)\otimes\bm\to\R$: if
\[
\begin{split}
A&=\sum_{j<i}a_{ij}(e_{ij}+e_{ji})+\sum_{i=1}^{n}a_{ii}e_{ii},\\
B&=\sum_{1\le j\le i\le n}b_{ij}e_{ij},
\end{split}
\]
where $\sum a_{ii}=0=\sum b_{ii}$, then
\[
\lt A,B\rt=\sum_{1\le j\le i\le n}a_{ij}b_{ij}.
\]
Thus this pairing allows one to identify the space $Symm_n(\R)$ with $\bm^*$; we will denote the corresponding map by $\phi:\sy\cong\bm^*$; it consists of ``deleting'' the strictly upper triangular part of a matrix in \sy\ while in the lower triangular part we raise all the indices. We can now use this identification to transfer the Poisson structure \eqref{pois1} from $\bm^*$ onto \sy. Let us write the explicit formulas for the induced Poisson structure. Let $f=f(a_{ij})\in C^\infty(\sy)$ and $L=(a_{ij})\in\sy,\,i\le j$ that is
\[
L=\sum_{1\le j<i\le n}a_{ij}(e_{ij}+e_{ji})+\sum_{i=1}^na_{ii}e_{ii}.
\]
We can use $\phi$ to induce from $f$ a function $f^\phi$ on $\bm^*$ simply by replacing every $a_{ij}$ by the coordinate $b^{ij}$ on the dual space $\bm^*$. Then $df^\phi(\phi(L))\in\bm$ can be identified with the lower triangular matrix of partial derivatives
\[
\sum_{j\le i}\frac{\dd f^\phi}{\dd b^{ij}}(\phi(L))e_{ij}=\sum_{j\le i}\frac{\dd f}{\dd a_{ij}}(L)e_{ij}\in\bm,
\]
so that we can use the formula \eqref{pois1}. However, for future applications it is more convenient to write everything in terms of ``gradients''.

For this we observe that since the pairing \eqref{pois1.5} is positive-definite on $\sy$, we can use it to identify $\sy^*\cong\sy$. This identification sends the elements $(e_{ij}+e_{ji})^*,\,j<i$ in $\sy^*$ dual to $e_{ij}+e_{ji}\in\sy$ into $\frac12(e_{ij}+e_{ji})$. Thus, if $f=f(a_{ij})\in C^\infty(\sy)$ and $L\in\sy$, then the differential of $f$ at $L$ is
\[
df(L)=\sum_{1\le j<i\le n}\frac{\dd f}{\dd a_{ij}}(L)(e_{ij}+e_{ji})^*+\sum_{i=1}^n\frac{\dd f}{\dd a_{ii}}e_{ii}^*,
\]
and its ``gradient'' is
\[
\nabla^s f(L)=\frac12\sum_{1\le j<i\le n}\frac{\dd f}{\dd a_{ij}}(L)(e_{ij}+e_{ji})+\sum_{i=1}^n\frac{\dd f}{\dd a_{ii}}e_{ii}.
\]
We use the superscript $s$ to clarify that we consider the gradient in the sense of the space \sy. It is easy to see now that
\[
df^\phi(\phi(L))=\sum_{j\le i}\frac{\dd f}{\dd a_{ij}}(L)e_{ij}=\bar M(\nabla^s f(L)),
\]
where $\bar M:\sy\to\bm$ is given by the formula
\[
\bar M(L)=2L_- -L_0;
\]
here we denote by $L_-$ the lower-triangular part of the matrix, and by $L_0$ its diagonal. Thus we come up with the following formula: for any $f,g\in C^\infty(\sy)$ and any $L\in\sy$
\beq{pois2.1}
\{f,g\}(L)=\lt L,[\bar M(\nabla^s f(L)),\bar M(\nabla^s g(L))]\rt.
\eeq
Observe that $\bar M$ is the restriction to \sy\ of the natural projection onto \bm\ in the direct sum decomposition:
\[
\slnr=\sonr\oplus\bm.
\]
Indeed for any $x=\sum_{i,j=1}^nx_{ij}e_{ij}$ we have
\[
x=\left(\sum_{1\le j<i\le n}(x_{ij}+x_{ji})e_{ij}+\sum_{i=1}^nx_{ii}e_{ii}\right)+\sum_{1\le j<i\le n}x_{ij}(e_{ij}-e_{ji}),
\]
where the first term on the right is just $\bar M(x)$, the second term is from \sonr, and $x$ is symmetric iff $x_{ij}=x_{ji}$. Let now $M=1-\bar M:\slnr\to\sonr$ be complement natural projection. If $L\in\sy$ is symmetric, then
\[
M(L)=L_+-L_-,
\]
the difference between its upper- and lower-triangular parts, i.e
\[
M\left(\sum_{j<i}a_{ij}(e_{ij}+e_{ji})+\sum_{i=1}^{n}a_{ii}e_{ii}\right)=\sum_{j<i}a_{ij}(e_{ij}-e_{ji}).
\]
Then for $f,g\in C^\infty(\sy)$ we have
\beq{pois3}
\begin{split}
\lt L,[\bar M(\nabla^sf(L)),&\bar M(\nabla^sg(L))]\rt=\lt L,[\nabla^sf(L),\nabla^sg(L)]\rt-\lt L,[\nabla^sf(L),M(\nabla^sg(L))]\rt\\
       &\quad-\lt L,[M(\nabla^sf(L)),\nabla^sg(L)]\rt+\lt L,[M(\nabla^sf(L)),M(\nabla^sg(L))]\rt\\
       &=\lt L,[M(\nabla^sg(L)),\nabla^sf(L)]\rt-\lt L,[M(\nabla^sf(L)),\nabla^sg(L)]\rt.
\end{split}
\eeq
This follows from the fact that \sy\ is orthogonal to \sonr\ with respect to the Killing pairing.

\begin{rem}\label{rempoi}
It is very useful to write down the Poisson structure on \sy\ in terms of the usual ``gradients'' of the functions, i.e. to the gradients, inherited from the Killing form on \slnr, rather then from its restriction to \sy. Of course, this can only work for functions $f\in C^\infty(\sy)$ that are restrictions of some functions on \slnr.

So let us consider an arbitrary functions $f\in C^\infty(\slnr)$, which we restrict to a function $f^s$ on \sy, and let $L\in\sy$ be an arbitrary point. Observe that $\nabla f(L)\neq\nabla^s f^s$ as matrices, in particular, the former matrix is not necessarily symmetric, unless $f(x^t)=f(x)$ for all $x\in\slnr$. However it is easy to see that by chain rule
\[
\frac{\dd f^s}{\dd a_{ij}}=\sum_{p,q=1}^n\frac{\dd f}{\dd x_{pq}}\frac{\dd x_{pq}}{\dd a_{ij}}=\frac{\dd f}{\dd x_{ij}}+\frac{\dd f}{\dd x_{ji}},
\]
so
\[
\nabla^s f^s(L)=\frac12\sum_{1\le j\le i\le n}\frac{\dd f^s(L)}{\dd a_{ij}}(e_{ij}+e_{ji})=\frac12\sum_{1\le j\le i\le n}\left(\frac{\dd f}{\dd x_{ij}}(L)+\frac{\dd f}{\dd x_{ji}}(L)\right)(e_{ij}+e_{ji})
\]
and therefore
\[
\bar M(\nabla^s f^s(L))=\sum_{1\le j<i\le n}\left(\frac{\dd f}{\dd x_{ij}}(L)+\frac{\dd f}{\dd x_{ji}}(L)\right)e_{ij}+\sum_{i=1}^n\frac{\dd f}{\dd x_{ii}}(L)e_{ii}=\bar M(\nabla f(L)).
\]
Hence for any two functions $f,g\in C^\infty(\mathfrak{sl}_n(\R))$ which we restrict onto $\sy$, their Poisson bracket is given by the formula
\beq{pois2.2}
\{f^s,g^s\}(L)=\lt L,[\bar M(\nabla f(L)),\bar M(\nabla g(L))]\rt
\eeq
for any $L\in\sy$.
\end{rem}

\subsection{The bracket of \Bp-invariant functions}\label{sect2.2}
From now on we will assume that $f,g$ are equal to restrictions onto \sy\ of some smooth functions on \slnr. Henceforth we will not distinguish between functions $f$ on \slnr\ and their restrictions to \sy, $f^s$, in our notation, unless we need to stress the difference. Due to the remark \ref{rempoi} we can substitute the gradients of $\nabla f,\nabla g$ of functions $f,g$ as functions on \slnr\ for the gradients $\nabla^s f^s,\nabla^s g^s$ of restrictions $f^s,g^s$ of $f,g$ onto \sy\ in the formula \eqref{pois2.1}: in fact the formulas \eqref{pois2.1} and \eqref{pois2.2} coincide in this case. The purpose of this move is pretty clear: the group $SL_n(\R)$ acts on its Lie algebra by conjugations and so do all its subgroups, while the conjugation by a matrix $g\in SL_n(\R)$ does not preserve the subspace \sy, unless $g$ is orthogonal.

The main result of this section is the following smooth version of the well-known Adler-Kostant-Symes scheme (which is usually proved in the context of polynomial functions):
\begin{prop}\label{instrumental}
Let $f,g\in C^\infty(\slnr)$ be \Bp-invariant functions, which we restrict to \sy, then for all $L\in\sy$ we have
\[
\{f^s,g^s\}(L)=-\lt L,[\nabla f(L),\nabla g(L)]\rt
\]
\end{prop}
This statement is based on the following lemma:
\begin{lemma}\label{lem1}
Let $f\in C^\infty(\slnr)$ be any function and $\xi\in\slnr$ be a vector. Then the following is true:
\[
\frac{\dd f}{\dd\xi}(x)=\lt \nabla f(x),[\xi,x]\rt,\ \mbox{for all}\ x\in\slnr,
\]
where $\frac{\dd f}{\dd\xi}\in C^\infty(\slnr)$ is the directional derivative of $f$ with respect to the vector field $ad(\xi)$ on $\slnr$, i.e.
\[
\frac{\dd f}{\dd\xi}(x)=\frac{df(Ad_{\exp(t\xi)}x)}{dt}|_{t=0}.
\]
In particular, if $f$ is invariant with respect to the field $ad(\xi)$, then
\[
-\lt x,[\nabla f(x),\xi^t]\rt=\lt \nabla f(x),[\xi,x]\rt=\frac{\dd f}{\dd\xi}(x)=0.
\]
\end{lemma}
Observe that in coordinates one can write down the vector field $\frac{\dd}{\dd\xi}$ as
\[
\dot x=[\xi,x]
\]
for a generic point $x\in\slnr$; this follows directly from the formula for $\frac{\dd f}{\dd\xi}$. Or else one can regard it as the corollary of the definition and the fact that the adjoint representation $ad$ on Lie algebras is equal to the linear part of the ``big'' adjoint representation $Ad$ of the group.

Let us first derive the statement of proposition \ref{instrumental} from the lemma:
\begin{proof}
For any two functions $f,g\in C^\infty(\slnr)$, restricted to \sy, we compute:
\beq{pois4}
\begin{split}
\{f^s,g^s\}(L)&=\lt L,[\bar M(\nabla f(L)),\bar M(\nabla g(L))]\rt\\
             &=\lt L,[(1-M)(\nabla f(L)),(1- M)(\nabla g(L))]\rt\\
             &=\lt L,[\nabla f(L),\nabla g(L)]\rt-\lt L, [M(\nabla f(L)),\nabla g(L)]\rt\\
             &\quad-\lt L,[\nabla f(L),M(\nabla g(L))]\rt+\lt L, [M(\nabla f(L)),M(\nabla g(L))]\rt\\
             &=\!\lt L,\![\nabla f(L),\!\nabla g(L)]\rt\!-\!\lt L,\![M(\nabla f(L)),\!\nabla g(L)]\rt\!-\!\lt L,\![\nabla f(L),\!M(\nabla g(L))]\rt.
\end{split}
\eeq
The last term disappears because $\sy\perp\sonr$. Let now $f,g\in C^\infty(\slnr)$ be \bp-invariant functions, then for all $L\in\sy$ we have from the formula \eqref{pois4}
\[
\begin{split}
\{f^s,g^s\}(L)&=\lt L,[\nabla f(L),\nabla g(L)]\rt-\lt L, [M(\nabla f(L)),\nabla g(L)]\rt-\lt L,[\nabla f(L),M(\nabla g(L))]\rt\\
             &=\!\lt L,\![\nabla\! f(L),\!\nabla\! g(L)]\rt\!-\!\lt L,\![(1\!-\!\bar M)(\nabla\! f(L)),\!\nabla\! g(L)]\rt\!-\!\lt L,\![\nabla\! f(L),\!(1\!-\!\bar M)(\nabla\! g(L))]\rt\\
             &=-\lt L,[\nabla f(L),\nabla g(L)]\rt+\lt L,[\bar M(\nabla f(L)),\nabla g(L)]\rt+\lt L,[\nabla f(L),\bar M(\nabla g(L))]\rt\\
             &=-\lt L,[\nabla f(L),\nabla g(L)]\rt,
\end{split}
\]
where the last two terms vanish because the functions $f,g$ are \Bp-invariant (and hence \bp-invariant), and $\bar M(\nabla f),\bar M(\nabla g)\in\bp$ see lemma \ref{lem1}.
\end{proof}
It now remains to prove lemma \ref{lem1}. To this end we recall that $Ad_gx=gxg^{-1}$ for all $g\in SL_n(\R),\,x\in\slnr$, so taking $g=g(t)=\exp(t\xi)$ we compute:
\[
\frac{d}{dt}f(Ad_{g(t)}x)_{|_{t=0}}=\sum_{p,q=1}^n\left(\frac{\dd f}{\dd x_{pq}}(Ad_{g(t)}x)\,\frac{d(Ad_{g(t)}x)_{pq}}{dt}\right)_{|_{t=0}}=\sum_{p,q=1}^n\frac{\dd f(x)}{\dd x_{pq}}[\xi,x]_{pq}=\lt\nabla f(x),[\xi,x]\rt.
\]
The second equation in the lemma \ref{lem1} is just the property of the Killing pairing: for any three matrices $A,B$ and $C$ we have
\[
\begin{split}
\lt A,[B,C]\rt&=\tr(A^t[B,C])=\tr(A^tBC)-\tr(A^tCB)\\&=\tr((CA^t-A^tC)B)=\tr([A,C^t]^tB)=\lt B,[A,C^t]\rt.
\end{split}
\]
Then if we put $C^t=\bar M(\nabla g(L))=\xi,\,A=\nabla f,\,B=L$, we get by the previous formula
\[
\lt L,[\nabla f(L),\bar M(\nabla g(L))]\rt=\lt \nabla f,[L,\bar M(\nabla g(L))^t]\rt=-\frac{\dd f}{\dd\xi}(L)=0,
\]
since $\xi=\bar M(\nabla g(L))^t\in\bp$ and $f$ is \Bp-invariant.
\begin{rem}\label{rem1imp}
Once again we draw the reader's attention to the fact that the formula from proposition \ref{instrumental} cannot be pulled to the functions defined only on \sy\ and their symmetric or non-symmetric gradients: as we observed before, the action of \Bp\ is not defined on \sy.

We also observe that one can regard the claim of the proposition \ref{instrumental} as the following statement: \textit{restriction of \Bp-invariant functions from \slnr\ to \sy\ is an anti-homomorphism of Poisson algebras:
\beq{pois5}
\{f,g\}^s=-\{f^s,g^s\}
\eeq
for all $f,g\in C^\infty(\slnr)^{\Bm}$.} In particular, it sends commutative families of functions to commutative, which is a version of the classical Adler-Kostant-Symes construction for smooth functions. Recall that the claim of the classical AKS scheme is as follows: \textit{let $\mathfrak g=\mathfrak g_1\oplus\mathfrak g_2$ be a direct sum decomposition of a Lie algebra as a sum of two Lie subalgebras and let $S\mathfrak g=S\mathfrak g_1\oplus \mathfrak g_2S\mathfrak g,\,f=f_1+f_2$ be the corresponding decomposition of the polynomial functions on $\mathfrak g^*$. Let $f,g\in S\mathfrak g$ be two Poisson-commuting functions, each of which is $\mathfrak g_1$-invariant. Then their restrictions to $\mathfrak g_1$ Poisson-commute with each other.}

One can modify the proof of this claim so as to obtain a statement, similar to that of the proposition \ref{instrumental} for polynomial functions; see for instance the electronic version of \cite{TCS} and the proof therein, which is based on the properties of the polynomials. On the other hand we couldn't find in literature a version of this statement for non-polynomial functions; thus we found it necessary to give here a detailed proof thereof.
\end{rem}
\begin{rem}\label{remhamfield}
One can use the formulas proved in this section to obtain the expression for the Hamilton field of a \Bp-invariant function restricted to \sy. Namely recall that the Hamilton field $X_f$ of a function $f$ is determined by the equation
\[
\{f,g\}=X_f(g)
\]
for any smooth function $g$ on a Poisson manifold. Let now $f$ be a \Bp-invariant function, and $g$ arbitrary function. Consider now the formula \eqref{pois4} and let $f$ be a \Bp-invariant function. Then, since $M+\bar M=1$, we have
\[
\begin{split}
\{f^s,g^s\}(L)&=\!\lt L,\![\nabla f(L),\!\nabla g(L)]\rt\!-\!\lt L,\![M(\nabla f(L)),\!\nabla g(L)]\rt\!-\!\lt L,\![\nabla f(L),\!M(\nabla g(L))]\rt\\
               &=\lt L,[\nabla f(L),\bar M(\nabla g(L))]\rt-\lt L,[M(\nabla f(L)),\nabla g(L)]\rt\\
               &=-\lt L,[M(\nabla f(L)),\nabla g(L)]\rt=-\lt\nabla g(L),[L,M(\nabla f(L)]\rt\\
               &=\frac{\partial g(L)}{\partial M(\nabla f(L))}.
\end{split}
\]
Since this is true for any $g\in C^\infty(\sy)$ (as any smooth function on a hyperplane can be extended to the whole space) we conclude that the Hamilton field of $f^s$ (the restriction of a \Bp-invariant function to \sy) is equal to $[M(\nabla f(L)),L]$; in other words the Hamilton equation corresponding to $f$ is
\[
\dot L=[M(\nabla f(L)),L].
\]
\end{rem}

\section{Vector fields on $SO_n(\R)$ and Poisson structures}
In this section we are going to investigate the relation of Poisson structures from the previous section and vector fields on the orthogonal group.
\subsection{Toda fields and their symmetries}
In our previous paper \cite{CSS23} we described the way one can construct symmetries of the full symmetric Toda system by similar constructions. Let us first recall the main ideas of this construction.

To this end recall that for a diagonal matrix $\Lambda$ we define the \textit{Toda field} $\tc^\Lambda$ on $SO_n(\R)$ by the formula:
\[
\tc^\Lambda(\Psi)=M(\Psi\Lambda\Psi^t)\Psi,
\]
where $\Psi\in SO_n(\R)$ is a generic point in the orthogonal group $SO_n(\R)$, i.e. $\Psi^t=\Psi^{-1}$. Our interest in studying such fields relies on the following fact:
\begin{prop}
Let $L_0\in\sy$ be a symmetric matrix, and $\Lambda$ a real diagonal matrix of its eigenvalues (we can write them in any fixed order). Let $\Psi_0\in SO_n(\R)$ be the orthogonal matrix for which $L_0=\Psi_0\Lambda\Psi_0^t$ (its existence is guaranteed by the general theory). Consider the solution $\Psi(t)$ of Cauchy problem:
\[
\dot\Psi(t)=\tc^\Lambda(\Psi(t)),\ \Psi(0)=\Psi_0.
\]
Then $L(t)=\Psi(t)\Lambda\Psi^t(t)$ is the solution of Cauchy problem
\beq{toda1}
\dot L=[M(L), L],\ L(0)=L_0.
\eeq
Equation \eqref{toda1} is called \textbf{full symmetric Toda system}.
\end{prop}
In \cite{CSS23} we modified the definition of the field $\tc^\Lambda$ as follows: let $X\in\slnr$ be an arbitrary matrix. Put
\[
\tc^X(\Psi)=M(X)\Psi,\ \Psi\in SO_n(\R).
\]
Then we proved the following formula
\[
[\tc^X,\tc^Y]=\tc^{[X,Y]}.
\]
In other words, the map
\[
\begin{split}
\tc:\slnr&\to Vect(SO_n(\R)),\\
X&\mapsto \tc^X
\end{split}
\]
is a representation of the Lie algebra \slnr\ in vector fields on the compact group $SO_n(\R)$. In the cited paper we used another construction to obtain suitable coefficient functions $f_X$ (for certain $X\in\slnr$) such that $f_X\tc^X$ begin to commute with $\tc^\Lambda$. This gives a large family of symmetries of the Toda system, in particular it allows one to show that it is integrable by Lie-Bianchi theorem (we will publish this result later).

\subsection{The map \tc}\label{sect3.2}
In this paper we are going to look in another direction: we are going to establish a relation between the structure of the algebras of vector fields on $SO_n(\R)$ and the Poisson-Lie algebras of functions on \slnr. To this end we will modify the map \tc\ as follows.

Let $\Lambda$ be a diagonal matrix, and $f\in C^\infty(\slnr)$. We put
\beq{ftoda1}
\tc^{f,\Lambda}(\Psi)=M(\nabla f(\Psi\Lambda\Psi^t))\Psi,\ \Psi\in SO_n(\R).
\eeq
Observe that here we use the usual ``gradient matrix'' $\nabla f$ rather then the ``symmetric gradient matrix'' $\nabla^sf$ of the restriction $f^s$ of $f$ to the subspace \sy\ even though the matrix $L=\Psi\Lambda\Psi^t$ is symmetric. In what follows the matrix $\Lambda$ will be fixed and we will omit it from our notation, i.e. we will write $\tc^f=\tc^{f,\Lambda}$.

Our purpose is to use the map $f\mapsto\tc^f$ as a way to relate the Poisson structure on \slnr\ and the vector fields on $SO_n(\R)$. To this end we will need the following facts about the maps involved in its construction.
\begin{enumerate}
\item[(\textit{i})] The projection $M$ verifies the ``Nijenhuis'' equation :
\[
M^2([X,Y])-M([M(X),Y]+[X,M(Y)])+[M(X),M(Y)]=0;
\]
for all $X,Y\in\slnr$; in particular if $X,Y\in\sy$ then
\[
[X,Y]=M([M(X),Y]+[X,M(Y)])-[M(X),M(Y)],
\]
since $M^2=M=\id$ on \sonr.
\item[(\textit{ii})] Let $\xi,\,\eta$ be two smooth vector fields on a Lie group $G$ (for $G=\Slnr$ or \Sonr\ in our case), given by the formula
\[
\xi(g)=dR_g(x(g)),\ \eta(g)=dR_g(y(g)),\ g\in G,
\]
where $x,y:G\to\mathfrak g$ be two smooth functions on $G$ with values in its Lie algebra and $R_g:G\to G$ denotes the right translation by $g\in G$; then the commutator of $\xi,\,\eta$ is equal to the vector field $\zeta$ on $G$, determined by the smooth function $z:G\to\mathfrak g$:
\[
z(g)=\lc_\xi y(g)-\lc_\eta x(g)-[x(g),y(g)].
\]
\end{enumerate}
Here $\lc_\xi y(g)$ denotes the derivative of vector-valued function ($\mathfrak g$-valued function) $y(g)$ along the vector field $\xi$ and similarly $\lc_\eta x(g)$.

We are going to compute the commutator of fields $\tc^f,\tc^g$. First, we observe that
\[
\tc^f(\Psi)=dR_\Psi(x(\Psi)),\ \mbox{for}\ x(\Psi)=M(\nabla f(\Psi\Lambda\Psi^t)).
\]
So we begin with computing $\lc_\eta x(\Psi_0)$ for the right-invariant vector field $\eta(\Psi)=dR_\Psi(Y)$ for $Y=M(\nabla g(\Psi_0\Lambda\Psi_0^t))$ (it is sufficient to consider right-invariant fields since the derivative of a vector-valued function with respect to a vector field is linear over functions $\lc_{f\xi+g\eta}(x)=f\lc_\xi x+g\lc_\eta x$): let $L_0=\Psi_0\Lambda\Psi_0^t$ then
\[
\nabla f((1+\epsilon Y)\Psi_0\Lambda\Psi_0^t(1-\epsilon Y))=\nabla f(L_0))+\epsilon\nabla\left(\frac{\dd f}{\dd Y}\right)(L_0)+\bar o(\epsilon),
\]
where we use the notation from lemma \ref{lem1}. So, using the conclusion of this lemma and skew-symmetry of $Y$ we get
\[
\lc_YM(\nabla f(\Psi\Lambda\Psi^t))(\Psi_0)=M\left(\nabla\left(\frac{\dd f}{\dd Y}\right)(L_0)\right)=M(\nabla(\lt x,[\nabla f(x),Y]\rt)(L_0)).
\]
In this formula $x$ is a generic point in \slnr, with respect to which we calculate the ``gradient matrix''. So we compute further:
\[
\nabla\lt x,[\nabla f(x),Y]\rt=[\nabla f(x),Y]+\lt x_{(1)},[\nabla^2 f(x),Y_{(1)}]\rt_{(1)}.
\]
Here $\nabla^2f(x)$ is the ``matrix of second partial derivatives'' of the function $f$: it is an element of the tensor square of \slnr, given by the formula (we again mutely use the Killing metric to raise the indices):
\[
\nabla^2 f(x)=\sum_{1\le i,j,k,l\le n}\frac{\dd^2 f(x)}{\dd x_{ij}\dd x_{kl}}e_{ij}\otimes e_{kl}.
\]
Further, $x_{(1)}=x\otimes\mathbbm 1, Y_{(1)}=Y\otimes\mathbbm 1$ and $\lt,\rt_{(1)}$ denotes the ``Killing pairing in the first coordinate'', i.e.
\[
\lt A,B\rt_{(1)}=\sum_{1\le i,j,k,l,m\le n}a_{ijkl}b_{ijlm}e_{km}
\]
if $A=\sum_{i,j,k,l}a_{ijkl}e_{ij}\otimes e_{kl},\,B=\sum_{i,j,k,l}b_{ijkl}e_{ij}\otimes e_{kl}$. Summing up, we get:
\[
\lc_YM(\nabla f(\Psi\Lambda\Psi^t))(\Psi_0)=M([\nabla f(L_0),Y])+M(\lt (L_0)_{(1)},[\nabla^2f(L_0),Y_{(1)}]\rt_{(1)}).
\]
Let us modify the second term on the right. To this end we put
\[
L_0=A=\sum_{i,j}a_{ij}e_{ij},\ Y=B=\sum_{k,l}b_{kl}e_{kl},\, \nabla^2f(L_0)=C=\sum_{p,q,r,s}c_{pqrs}e_{pq}\otimes e_{rs}
\]
where $A$ is symmetric, $a_{ij}=a_{ji}$ and $B$ is skew-symmetric, $b_{kl}=-b_{lk}$. Then $A^t_{(1)}=A_{(1)}$, where we apply transposition ``in the first tensor direction''. Let $\tr_1$ denote the trace ``in the first tensor direction'', it clearly verifies the same cyclic property and thus
\[
\lt A_{(1)},[C,B_{(1)}]\rt_{(1)}=\tr_1(A_{(1)}^tCB_{(1)}-A_{(1)}^tB_{(1)}C)=\tr_1([B_{(1)},A_{(1)}]C)=\lt[B,A]_{(1)},C\rt_{(1)},
\]
since the commutator of a symmetric and a skew-symmetric matrix is symmetric. So
\[
M(\lt (L_0)_{(1)},[\nabla^2f(L_0),Y_{(1)}]\rt_{(1)})=M(\lt [Y,L_0]_{(1)},\nabla^2f(L_0)\rt_{(1)})=\lt [Y,L_0]_{(1)},M_{(2)}(\nabla^2f(L_0))\rt_{(1)},
\]
where $M_{(2)}$ denotes the ``action of projector $M$ on the second tensor coordinate'':
\[
M_{(2)}\left(\sum_{p,q,r,s}c_{pqrs}e_{pq}\otimes e_{rs}\right)=\sum_{\substack{1\le p,q\le n\\ 1\le s<r\le n}}c_{pqrs}e_{pq}\otimes (e_{rs}-e_{sr}).
\]
So we can finally return $Y$ to the right
\[
M(\lt (L_0)_{(1)},[\nabla^2f(L_0),Y_{(1)}]\rt_{(1)})=\lt (L_0)_{(1)},[M_{(2)}(\nabla^2f(L_0)),Y_{(1)}]\rt_{(1)}
\]
Summing up, we get $[\tc^f,\tc^g](\Psi)=dR_\Psi(z(\Psi))$, where $z(\Psi)\in\sonr$ is given by the following formula: let $L=\Psi\Lambda\Psi^t$ for some real diagonal matrix $\Lambda$ then with the help of property (\textit{i}) of $M$ we obtain
\beq{commut1}
\begin{split}
z(\Psi)&=\,M([\nabla f(L),M(\nabla g(L))])+\lt L_{(1)},[M_{(2)}(\nabla^2f(L)),M(\nabla g(L))_{(1)}]\rt_{(1)}\\
          &\quad-M([\nabla g(L),M(\nabla f(L))])-\lt L_{(1)},[M_{(2)}(\nabla^2g(L)),M(\nabla f(L))_{(1)}]\rt_{(1)}\\
          &\qquad\qquad-[M(\nabla f(L)),M(\nabla g(L))]\\
          &=M([\nabla f(L),\nabla g(L)])+\lt L_{(1)},[M_{(2)}(\nabla^2f(L)),M(\nabla g(L))_{(1)}]\rt_{(1)}\\
          &\qquad\qquad+\lt L_{(1)},[M(\nabla f(L))_{(1)},M_{(2)}(\nabla^2g(L))]\rt_{(1)}
\end{split}
\eeq
\subsection{Fields $\tc^f$ for \Bp-invariant functions}\label{sect3.3}
Let $f,g\in C^\infty(\slnr)$ now be \Bp-invariant functions. We are going to show that in this case the following is true:
\begin{prop}
If $f,g\in C^\infty(\slnr)$ are \Bp-invariant, then
\[
[\tc^f,\tc^g]=\tc^{\{f,g\}},
\]
where the Poisson brackets in the right hand side signifies the Kirillov-Kostant structure on \slnr\ (see formulas \eqref{pois1},\,\eqref{pois1.5}).
\end{prop}
In other words, the map $\tc:C^\infty(\slnr)^{\Bp}\to Vect(\Sonr)$ is a homomorphism of Lie algebras. Observe that this is to be expected, as according to the remark \ref{remhamfield} the fields $\tc^f$ for generate Hamilton fields of the \Bp-invariant functions $f$, restricted to \sy: if $\Psi(t)$ is a trajectory of the field $\tc^f$, then the curve $L(t)=\Psi(t)\Lambda\Psi(t)^t$ verify the equation
\[
\dot L=[M(\nabla f(L)),L],
\]
as we need. Thus one is tempted to derive this formula from the well-known property of the Hamilton fields:
\[
[X_f,X_g]=X_{\{f,g\}}.
\]
However the relations between the fields on \Sonr\ and the functions on \sy\ is rather indirect, there can be kernels in both directions (especially if the matrix $L$ has non-simple spectrum), hence the commutation relation on \sy\ does not automatically entail similar formulas for \Sonr. So we find it necessary to prove this independently; it is interesting that this construction has very little to do with the symmetricity etc. of the matrices; in particular one can prove a similar formula for the fields induced by the formula
\[
\tc^{f,X}(\Psi)=M(\nabla f(\Psi X\Psi^t))\Psi,\ \Psi\in\Sonr,\ \mbox{for any}\ X\in\slnr.
\]
This, however would demand a long discussion, so we postpone it to a forthcoming paper.
\begin{proof}
Recall that $M,\bar M$ denote the projections onto \sonr\ and \bm\ respectively in the direct sum decomposition $\slnr=\sonr\oplus\bm$, so $M+\bar M=\id$ and $\bar M(X)\in\bm$ for all $X\in\slnr$. Observe that for \Bp-invariant functions $f,\,g$ one has the following formula:
\[
\lt x, [\nabla f(x),\bar M(\nabla g(x))]\rt=-\frac{\dd f(x)}{\dd \bar M(\nabla g(x))^t}=0=\frac{\dd g(x)}{\dd \bar M(\nabla f(x))^t}=\lt x,[\bar M(\nabla f(x)),\nabla g(x)]\rt.
\]
This follows directly from lemma \ref{lem1}. Now similarly we have
\[
\lt x_{(1)},[\nabla^2 f(x),\bar M(\nabla g(x))_{(1)}]\rt_{(1)}=-\frac{\dd(\nabla f)(x)}{\dd \bar M(\nabla g(x))^t}=-\nabla\left(\frac{\dd f(x)}{\dd \bar M(\nabla g(x))^t}\right)=0,
\]
where we regard $\nabla f$ as a matrix-valued function and use the commutativity of partial derivatives with respect to constant vectors. For the same reason we have
\[
\lt x_{(1)},[\bar M(\nabla f(x))_{(1)},\nabla^2 g(x)]\rt_{(1)}=0.
\]
It follows that since $M+\bar M=\id$
\beq{equal}
\begin{split}
\lt x_{(1)},[\nabla^2 f(x),\nabla g(x)_{(1)}]\rt_{(1)}&=\lt x_{(1)},[\nabla^2 f(x),M(\nabla g(x))_{(1)}]\rt_{(1)},\\
\lt x_{(1)},[\nabla f(x)_{(1)},\nabla^2 g(x)]\rt_{(1)}&=\lt x_{(1)},[M(\nabla f(x))_{(1)},\nabla^2 g(x)]\rt_{(1)}.
\end{split}
\eeq
Finally, consider the field $\tc^{\{f,g\}}$: by definition
\[
\tc^{\{f,g\}}(\Psi)=M(\nabla\{f,g\}(L))\Psi,
\]
where $L=\Psi\Lambda\Psi^t$. First we compute $\nabla\{f,g\}(x)$: by formula \eqref{pois1.5} we have
\[
\begin{split}
\nabla\{f,g\}(x)&=\nabla(\lt x,[\nabla f(x),\nabla g(x)]\rt)=\\&=[\nabla f(x),\nabla g(x)]+\lt x_{(1)},[\nabla^2f(x),\nabla g(x)_{(1)}]\rt_{(1)}+\lt x_{(1)},[\nabla f(x)_{(1)},\nabla^2g(x)]\rt_{(1)}
\end{split}
\]
Hence by equations \eqref{equal}
\[
\begin{split}
\nabla&\{f,g\}(x)=\\
          &=[\nabla f(x),\nabla g(x)]+\lt x_{(1)},[\nabla^2f(x),M(\nabla g(x))_{(1)}]\rt_{(1)}+\lt x_{(1)},[M(\nabla f(x))_{(1)},\nabla^2g(x)]\rt_{(1)}.
\end{split}
\]
Now the claim follows by comparing this formula with equation \eqref{commut1}.
\end{proof}
\begin{rem}
A simple corollary of this claim is that there exist large commutative families of fields commuting with Toda fields $\tc^\Lambda$; in effect, any field $\tc^f$ for \Bp-invariant $f$ will commute with $\tc^\Lambda$. To this end it is sufficient to observe that $\tc^\Lambda=\tc^{\frac12\tr(x^2)}$, i.e. $\tc^\Lambda$ is the field associated with the function $g=\frac12\tr(x^2)\in C^\infty(\slnr)$, and this function is central with respect to the Poisson structure \eqref{pois1},\,\eqref{pois1.5}.
\end{rem}
\section{Examples and further discussion}
In this section we suggest few examples of the phenomena, we discussed earlier: we begin with an example of \bp-invariant functions on \slnr\ (for $n=4$), and calculate explicitly the corresponding Hamilton vector fields ($M$-operators). After this we discuss the relation between $M$-operators and the vector fields on the orthogonal group.
\subsection{Lax representation for \Bp-invariant functions and vector fields}
As we described earlier (see remark \ref{remhamfield}), the Hamilton equation, associated with a \Bp-invariant function restricted to \sy, has form
\[
\dot L=[M(\nabla f(L)),L].
\]
Here the operator $\nabla f$ is the \slnr-valued ``gradient matrix'' of $f$ and $M$ is the natural projection onto \sonr\ along \bm. This is what one calls ``$M$-operator'' equation, which plays an important role in the theory. Let us give an example of calculations of such $M$-operators for particular choice of functions. Our exposition here is based on numerous discussions with D.Talalaev.
\begin{ex}[Functions on $Symm_4(\R)$]
Let us consider the full symmetric Toda system on symmetric matrices of size $4\times 4$;  
to this end we consider the following Lax matrix
\beq{L4-1}
L=A=
\begin{pmatrix}
 a_{11} & a_{12} & a_{13} & a_{14}\\
 a_{12} & a_{22} & a_{23} & a_{24}\\
 a_{13} & a_{23} & a_{33} & a_{34}\\
 a_{14} & a_{24} & a_{34} & a_{44}
\end{pmatrix}.
\eeq
The dimension of the phase space of this system is equal to $8$, which is equal to the number of independent variables (which is $10$) minus the number of Casimirs (which there are $2$). This system is superintegrable and it has five integrals of motion (Hamiltonians), see \cite{CS2}, where all these integrals are explicitly written down. Three of these integrals of motion are ``isospectral'', i.e. arise from the invariants of the Lax matrix: $\frac{1}{2}\tr(L^{2}),\,\frac{1}{3}\tr(L^{3}),\,\frac{1}{4}\tr(L^{4})$. The remaining integral of motion is given by the formula
\[
I = \frac{a_{41}^{(3)}}{a_{41}},
\]
where $a_{41}^{(3)}$ denotes the $(4,1)$-th matrix element of $L^{3}=L \cdot L \cdot L$. This integral of motion is equivalent to the integral
\[
I_1=\frac{A_{\frac{234}{123}}}{a_{14}}
\]
obtained by the chopping procedure (see \cite{DLNT}). Here $A_{\frac{234}{123}}$ denotes the minor obtained by deleting the first row and the last column of the Lax matrix. It turns out that this system has an additional integral of motion:
\[
J=\frac{A^{(2)}_{\frac{34}{12}}}{A_{\frac{34}{12}}},
\]
where $A_{\frac{34}{12}}$ is the determinant of the $2 \times 2$ submatrix of $L$ spanned by the first two columns and the last two rows of $A$ and $A^{(2)}_{\frac{34}{12}}$ denotes the corresponding $2 \times 2$ minor of $A^{2}$. Note that the functions $I$ and $J$ are equal to the restrictions of \Bp-invariant functions onto \sy: these functions are given by the same formulas as those for $I$ and $J$, but without restricting the attention to symmetric matrices.

By the formulas we proved earlier, in order to get the $M$-operators corresponding to functions $I$ and $J$ we should regard them as the functions on $\mathfrak{sl}_{4}(\mathbb{R})$, calculate their \slnr-gradients and anti-symmetrize the results. Here and below for the sake of brevity we denote functions on $\mathfrak{sl}_{4}(\mathbb{R})$ by the same symbols we use for functions on $Symm_4(\R)$. So let us start with the function:
\[
I = \frac{x_{41}^{(3)}}{x_{41}},
\]
where $x_{41}^{(3)}$ is the $(4,1)$-th matrix element of $X^{3}=X \cdot X \cdot X,\,X \in \mathfrak{sl}_{4}(\mathbb{R})$. It turns out that it is easier to calculate the \slnr-gradient for the function $\tilde{I}=I-\frac{1}{2}\tr(X^{2})$ and project it into $\sof$ along \bmf:
\[
M(\nabla \tilde{I})= M\left(\nabla\left(I-\frac{1}{2}\tr(X^{2}\right)\right).
\]
In fact, we can use the linearity of this procedure to simplify some of the computations: in effect we know that the $M$-operator for $\frac{1}{2}\tr(X^{2})$ is equal to $M=L_{>0}-L_{<0}$ for all $X=L \in\sy$. First of all we calculate \slnr-gradient: put
\[
F_{ab} = \frac{\partial}{\partial x_{ab}}\left(I-\frac{1}{2}\tr(X^{2})\right),
\]
where
\[
 I=\frac{x_{4i}x_{ij}x_{j1}}{x_{41}},\quad \frac{1}{2}\tr(X^{2})=\frac{1}{2}x_{kl}x_{lk},
\]
and summation upon repeating indices in the range from $1$ to $4$ is understood. Then
\beq{equal}
F_{ab} = \frac{1}{x^{2}_{41}}\left(\frac{\partial\left(x_{4i}x_{ij}x_{j1}-\frac{1}{2}x_{kl}x_{lk}x_{41}\right)}{\partial x_{ab}}x_{41} - \frac{\partial x_{41}}{\partial x_{ab}}\left(x_{4i}x_{ij}x_{j1}-\frac{1}{2}x_{kl}x_{lk}x_{41}\right)\right).
\eeq
We are interested only in matrix elements $F_{ab}$ with $a < b$, because we will project along \bmf into \sof\ afterwards, so let
\[
F=
\begin{pmatrix}
 \ast & F_{12} & F_{13} & F_{14}\\
 \ast & \ast & F_{23} & F_{24}\\
 \ast & \ast & \ast & F_{34}\\
 \ast & \ast & \ast & \ast
\end{pmatrix}.
\]
As $\frac{\partial x_{41}}{\partial x_{ab}} = 0$ for all $a < b$, we get
\[
\begin{split}
&\frac{\partial\left(x_{4i}x_{ij}x_{j1}-\frac{1}{2}x_{kl}x_{lk}x_{41}\right)}{\partial x_{ab}}=\\
&\qquad\qquad=\delta_{4a}\delta_{ib}x_{ij}x_{j1} + x_{4i}\delta_{ia}\delta_{jb}x_{j1} + x_{4i}x_{ij}\delta_{ja}\delta_{1b} -  \frac{1}{2}\delta_{ka}\delta_{lb}x_{lk}x_{41} -  \frac{1}{2}x_{kl}\delta_{la}\delta_{kb}x_{41}\\
&\qquad\qquad=\delta_{4a}x_{bj}x_{j1} + x_{4a}x_{b1} + x_{4i}x_{ia}\delta_{1b} - x_{ba}x_{41}.
\end{split}
\]
We finally get
\[
F= \frac{1}{x_{41}}
\begin{pmatrix}
 \ast & 0 & 0 & 0\\
 \ast & \ast & X_{\frac{34}{12}} & 0\\
 \ast & \ast & \ast & 0\\
 \ast & \ast & \ast & \ast
\end{pmatrix}, \ \mbox{where}\ X_{\frac{34}{12}}=x_{31}x_{42}-x_{32}x_{41}.
\]
Eventually after projection into \sof\ along \bmf\ and substituting symmetric matrix: $x_{ij}=x_{ji}=a_{ij}, \ i < j$ we get
\[
M(\nabla \tilde{I})= \frac{1}{a_{14}}
\begin{pmatrix}
 0 & 0 & 0 & 0\\
 0 & 0 & A_{\frac{34}{12}} & 0\\
 0 & -A_{\frac{34}{12}} & 0 & 0\\
 0 & 0 & 0 & 0
\end{pmatrix},\ \mbox{where}\ A_{\frac{34}{12}}=a_{13}a_{24}-a_{23}a_{14}.
\]
In a similar way we get for function $J$:
\[
M(\nabla J)=  \frac{1}{A_{\frac{34}{12}}}
\begin{pmatrix}
0 & -B_{1} & 0 & 0\\
B_{1} & 0 & 0 & 0\\
0 & 0 & 0 &  -B_{2}\\
0 & 0 & B_{2} & 0
\end{pmatrix},\ \mbox{where}\ B_{1}=a^{(2)}_{14}a_{13}-a^{(2)}_{13}a_{14}, \ B_{2}=a^{(2)}_{14}a_{24}-a^{(2)}_{24}a_{14}.
\]
\end{ex}
One of the main purposes of this computation is to check the conclusions of our previous sections. For instance, direct calculations show that the commutators of vector fields $\tc^{\tilde{I}}, \tc^J$ and the fields $\tc^{\frac{1}{k}Tr L^{k}}$, corresponding to ``isospectral'' integrals of motion are equal to zero, and the commutator $[ \tc^{\tilde{I}}, \tc^J ]$ is not equal to zero. It is to be expected because the functions $I$ and $J$ Poisson-commute with the isospectral integrals and do not commute with each other.

In fact, it is easy to see that $[ \tc^{\tilde{I}}, \tc^J ] \neq 0$ even without calculation. Roughly speaking the commutator of the vector fields $\tc^{\tilde{I}}$ and $\tc^J$ is a linear combination of the derivatives $\tc^{\tilde{I}}(M(\nabla J))$, $\tc^J (M(\nabla \tilde{I}))$ of matrix-valued functions $M(\nabla J),\,M(\nabla \tilde{I})$ and the matrix commutator of these functions $[M(\nabla J), M(\nabla \tilde{I})]$.
It is clear that the latter commutator of matrices has nonzero elements in positions $(1,3), \ (2,4)$ and $(3,1), \ (4,2)$ and these elements cannot be eliminated by other terms 
because the expressions $\tc^{\tilde{I}} (M(\nabla J))$ and $\tc^J (M(\nabla \tilde{I}))$ can have nonzero elements only at the same positions where such elements appear in $M(\nabla J)$ and $M(\nabla \tilde{I})$.

\subsection{From $M$-operators on \sy to vector fields on \Sonr}
In our previous sections we described vector fields on \Sonr, associated with \Bp-invariant functions. We remarked that if we try to do the ``inverse move'', i.e. if we restore the fields on \sy\ from dynamics on \Sonr\ (induce them from the adjoint action of the group \Sonr\ on the space of symmetric matrices), then we obtain the Hamilton fields of the original functions.

It turns out that it is possible to some extent to restore the fields on \Sonr\ from the fields on \sy; more accurately one can show that if fields have the prescribed special form (the $M$-operator form), then they are induced by the field of the form $\tc^f$ on \Sonr. Indeed, 
let us start with symmetric Lax matrix $L$; choose its polar decomposition:
\[
L = \Psi \Lambda \Psi^{t}, \ \Psi \in SO_n(\R).
\]
Suppose that a vector field on \sy\ has the form of an $M$-operator, i.e.:
\[
\tilde\tc^{f}(L) = [M(\nabla f(L)),L] = M(\nabla f(L))L - LM(\nabla f(L)),
\]
for some \Bp-invariant function. On the other hand, substituting the polar decomposition we may assume (locally) that there exists some vector field $\hat\tc^f$ on \Sonr, which by conjugation on symmetric matrices induces the dynamics of $\tilde\tc^f$. Then for this vector field we have:
\[
\tilde\tc^{f}(L) = \tilde\tc^{f}(\Psi\Lambda\Psi^{t}) = \hat\tc^{f}(\Psi)\Lambda \Psi^{t} +  \Psi \Lambda\hat\tc^{f}(\Psi^{t}).
\]
Taking in account that for any vector field $\hat\tc^f$ its value on a constant matrix-valued function vanishes, we get
\[
0=\hat\tc^f(\mathbbm 1)=\hat\tc^{f}(\Psi \Psi^{t}) = \hat\tc^{f}(\Psi) \Psi^{t} +  \Psi\hat\tc^{f}(\Psi^{t}) = 0.
\]
Thus we obtain
\[
\hat\tc^{f}(\Psi^{t}) = - \Psi^{t}\hat\tc^{f}(\Psi)\Psi^{t},
\]
and finally
\[
\tilde\tc^{f} L = \hat\tc^{f}(\Psi)\Psi^{t}\Psi \Lambda \Psi^{t} -  \Psi\Lambda\Psi^{t}\hat\tc^{f}(\Psi)\Psi^{t}.
\]
Comparing with the formula of the Lax dynamic, we get: a possible solution of this equation is
\[
\hat\tc^{f}(\Psi) = M(\nabla f(L))\Psi =\tc^f(\Psi).
\]
Once again, we remark that the vector field $\tc^f$ is not in general a unique solution of the equation for $\hat\tc^f$; for instance, one can change the eigen-matrix $\Lambda$ for another diagonal matrix $\Lambda'$ (with different order of eigenvalues). The situation is even worse in the case when the spectrum of $L$ is not simple, which allows families of anti-symmetric matrices, commuting with it.

\section{Conclusions}
Let us end this paper by a short list of possible questions for future investigations; we are not trying to exhaust the topic here, but merely give a list of qudirectly related with those we discuss here.

First of all, it goes without saying that all the statements and formulas discussed here (except for the explicit computations of matrix gradients and vector fields) have direct analogs for arbitrary Cartan pair, i.e for the full symmetric Toda systems, associated with arbitrary real semisimple Lie algebras. The actual equations are subject of future investigation, but seem to be direct generalizations of the results, present in this text.

Second, it is extremely interesting, how the vector fields $\tc^f$ we construct here interact with the symmetries of the full symmetric Toda field, found earlier in \cite{CSS23}. Both constructions are heavily relied on the use of the projector $M$, however in the latter case we need to multiply the results by suitable functions, which appear from the representation theory. Thus the question is far from being trivial.

Further, as one knows, full symmetric Toda system is related with the geometry of the flag space $\Slnr/\Bp=\Sonr/T_n$ (where $T_n$ is the intersection of the groups \Bp\ and \Sonr). It seems that the vector field $\tc^f$ for a \Bp\-invariant function $f$ can be unambiguously transferred onto the flag space; the role and significance of vector fields $\tc^f$ in this context is yet to be investigated.

Next, in addition to the full symmetric Toda system, one has another ``full'' generalisation of the Toda chain. We mean the famous \textit{full Kostant-Toda system}, when the role of symmetric matrices is played by another subspace in \slnr\ (this time an affine subspace). Whether similar equations or their modifications hold in that case is a subject for future investigations.

Finally, one can ask, for what decompositions $\mathfrak g=\mathfrak g_1\oplus\mathfrak g_2$ equalities similar to those, we consider in this paper, hold, and under what conditions. This might give rise to some new families of integrable systems on symmetric spaces.



\paragraph{Acknowledgements}
The work Yu.B.Chernyakov was carried out within the state assignment of NRC "Kurchatov institute". The work of G.Sharygin has been carried out during his visit to the Sino-Russian Mathematical center (Peking University) and supported by the National Key R\&D Program of China (Grant No. 2020YFE0204200). Both authors express their deepest gratitude to D.Talalaev for numerous fruitful discussions and valuable remarks.

\end{document}